\documentclass[reprint, aps, prl, english, showpacs]{revtex4-1}

\usepackage[T1]{fontenc}
\usepackage{amssymb}
\usepackage{graphicx}
\usepackage{amsmath,color}
\usepackage{natbib,hyperref}
\usepackage[squaren,Gray]{SIunits}
\usepackage{multirow,bigdelim}
\usepackage{amsfonts, bm}

\usepackage{lipsum}

\usepackage{color}

\begin{document}



\title{Correlations from ion-pairing and the Nernst-Einstein equation}

\author{Arthur France-Lanord}
\affiliation{Department of Materials Science and Engineering, Massachusetts Institute of Technology, Cambridge, Massachusetts 02139, USA}

\author{Jeffrey C. Grossman}
\affiliation{Department of Materials Science and Engineering, Massachusetts Institute of Technology, Cambridge, Massachusetts 02139, USA}

\email[]{jcg@mit.edu}

\date{\today}

\begin{abstract}
We present a new approximation to ionic conductivity well suited to dynamical atomic-scale simulations, based on the Nernst-Einstein equation. In our approximation, ionic aggregates constitute the elementary charge carriers, and are considered as non-interacting species. This approach conveniently captures the dominant effect of ion-ion correlations on conductivity, short range interactions in the form of clustering. In addition to providing better estimates to the conductivity at a lower computational cost than exact approaches, this new method allows to understand the physical mechanisms driving ion conduction in concentrated electrolytes. As an example, we consider Li$^+$ conduction in poly(ethylene oxide), a standard solid-state polymer electrolyte. Using our newly developed approach, we are able to reproduce recent experimental results reporting negative cation transference numbers at high salt concentrations, and to confirm that this effect can be caused by a large population of negatively charged clusters involving cations. 
\end{abstract}


\maketitle


Understanding ion transport is of fundamental importance in many different scientific domains, ranging from biophysics\cite{fulinski1997} to chemistry\cite{kennedy2016}. For instance, the design of new and better performing lithium-ion batteries\cite{bachman2015} requires us to be able to rapidly evaluate the ionic conductivity of a given electrolyte, as well as its cation transference number. If such a method were to be available, scientists would be able to computationally screen large databases of compounds, providing a complementary approach to experimental measurements, as well as a means to grasp fundamental design principles of electrolytes through the understanding of elementary transport mechanisms governing ion conduction. 

A simple way to tentatively increase the conductivity of an electrolyte is to increase its charge carrier concentration. However, in doing so, ionic interactions become more important\cite{qiao2018}, which leads to ion pairing and clustering -- effectively reducing the number of charge carriers. In fact, it is well known that some solid-state polymer electrolytes for instance show an optimal salt concentration for which conductivity is maximal\cite{lascaud1994}. Traditionally, the Nernst-Einstein equation is used to determine ionic conductivity and transference numbers from the ion's diffusion coefficients. However, for this expression to be exact, there need to be no interactions between ions whatsoever. As expected, this relation breaks down\cite{marcolongo2017,carbone2016,carbone2015} when ionic correlation becomes important, with experimental reports\cite{carbone2017} of discrepancies up to several orders of magnitude in conductivity: there is therefore a need for a reliable method taking into account ion-ion interactions. In this Letter, we develop an intermediate Nernst-Einstein approximation in terms of ionic clusters, which allows us to greatly reduce the error on the conductivity, at no additional computational cost. We evaluate an upper bound to the accuracy of this new approach using a non-interacting toy model. Finally, we apply our method to a realistic case study, Li$^+$ transport in poly(ethylene oxide), and confirm recent experimental observations\cite{pesko2017} of negative cation transference numbers at high salt concentrations. \\


In order to compute the ionic conductivity $\sigma$ of an electrolyte, one usually relies on the Nernst-Einstein equation, which states that the conductivity is proportional to the diffusion coefficients of the ionic species, and to their concentration. That is, for a $1:1$ salt: 

\begin{equation}\label{eq:NE}
\sigma_{\text{NE}} = \displaystyle\frac{e^2}{Vk_BT} \left( N_+ z_+^2 \bar{D}_+ + N_- z_-^2 \bar{D}_- \right), 
\end{equation}

where $e$ is the elementary charge, $k_B$ the Boltzmann constant, $V$ and $T$ respectively the volume and the temperature of the system of interest, and $\bar{D}_{\pm}$, $z_\pm$, $N_\pm$ respectively the diffusion coefficient, charge, and number of cations and anions. The diffusions coefficients are typically computed by monitoring the slope of the ion's mean squared displacement (MSD): 

\begin{equation}\label{eq: msd}
\bar{D}_{\pm} = \lim_{t' \to + \infty} \frac{\left\langle \left[ \mathbf{x}_i (t+t') - \mathbf{x}_i (t) \right]^2 \right\rangle_\pm}{6t'}, 
\end{equation}

where $\mathbf{x}_i(t)$ is the particle's instantaneous position, $t'$ the lag time, and $\left\langle \cdots \right\rangle$ denotes an ensemble average. An alternative but formally equivalent approach exists in the form of a Green-Kubo equation\cite{peng2016}. The ensemble average is obtained by averaging over time origins, usually using block averaging, and by averaging over all cations or anions. The diffusion coefficients obtained are comparable to what is probed using pulsed-field gradient nuclear magnetic resonance (PFG -- NMR)\cite{price1997}. \\

The Nernst-Einstein equation is based on the assumption that ions do not interact with themselves, and is therefore only exact in the infinite dilution limit. This approximation holds however rather well for dilute systems, especially when the dielectric constant of the solvent is high. For concentrated solutions, the assumption of non-interacting species breaks down, and failures of the Nernst-Einstein relation have been reported\cite{patro2016,chowdhuri2001,boden1991,carbone2016,carbone2015,carbone2017}. Ionic transport shifts from a single-particle to a many-particles picture. At high salt concentrations or with weak solvents, ions interact in the form of pairing, clustering, or at long-range; in other words, their motion is correlated. From linear response theory\cite{kubo1957}, an exact expression of the ionic conductivity accounting for correlations can be derived, here in an Einstein form: 

\begin{equation}\label{eq:Einstein}
\begin{split}
\sigma_{\text{E}} = &\lim_{t' \to + \infty}  \frac{e^2}{6t' Vk_BT} \sum_{ij} z_i z_j \\
& \left\langle \left[ \mathbf{x}_i (t+t') - \mathbf{x}_i (t) \right] \cdot
\left[ \mathbf{x}_j (t+t') - \mathbf{x}_j (t) \right] \right\rangle, 
\end{split}
\end{equation}

and equivalently using a Green-Kubo form\cite{french2011}. The Nernst-Einstein conductivity can be recovered by only considering diagonal ($i=j$) terms; all correlations arise from the off-diagonal terms. In equation \ref{eq:Einstein}, the central quantity is a \textit{collective} MSD. It is a property of the whole system itself, which means that contrary to the ions' MSD, the collective MSD cannot be averaged over particles, leading to an error on the conductivity at least larger by $\sqrt{N_{\pm}}$\cite{muller1995}. In practice, it is notoriously difficult to evaluate the true conductivity through this approach\cite{monteiro2008,taherkhani2017}. Significantly more sampling is required compared to the Nernst-Einstein conductivity. In addition, transference numbers can only be obtained when computing the whole matrix of binary transport coefficients as defined by Wheeler and Newman\cite{wheeler2004,newman2012} in the Stefan--Maxwell framework, whose off-diagonal matrix elements are strongly affected by noise. There is therefore a need for an intermediate approach, which would provide good estimates of the ionic conductivity in the correlated regime at a reasonable computational cost. Ideally, the method would also allow for a microscopic understanding of ionic correlations. \\

In the approach we present here, the main hypothesis is that ionic correlations mostly translate to short-range interactions, in the form of \textit{clustering}. Ionic clusters are known to form in electrolytes at high salt concentrations\cite{duluard2008}, which has the primary effect of reducing the conductivity, by lowering the number of free charge carriers through the formation of neutral clusters. It is common practice to scale down the Nernst-Einstein conductivity using a measure of the fraction of free charge carriers $\alpha$\cite{feng2018}; this correction however only captures one part of the picture, and a more general method can be derived. 

Consider an electrolytic system composed of solvent particles, anions, and cations. At a time $t$, its cluster population can be defined using a matrix $\bm{\alpha} (t)$, where the matrix element $\alpha_{ij}(t)$ corresponds to the number of clusters containing $i$ cations and $j$ anions. At thermodynamic equilibrium, the ensemble average $\left\langle \bm{\alpha} \right\rangle$ is time-independent; provided that the system of interest is ergodic, we can compute $\left\langle \bm{\alpha} \right\rangle$ by averaging over long-enough molecular dynamics simulations. In addition, ionic clusters do not diffuse at the same rate: a distinct diffusion coefficient $D_{ij}$ can be computed for each type of cluster. In matrix form, we can accordingly consider all cluster diffusion coefficients as components of a matrix $\bm{D}$, which has the same size and dimension as $\bm{\alpha}$. One can therefore reformulate the Nernst-Einstein equation to account for clustering: 

\begin{equation}\label{eq:cNE}
\sigma_{\text{cNE}} = \frac{e^2}{Vk_BT} \sum_{i=0}^{N_+}\sum_{j=0}^{N_-} z_{ij}^2 \alpha_{ij} D_{ij}, 
\end{equation}

where $z_{ij} = i z_+ + j z_-$ is the charge number of clusters, and $N_\pm$ is the number of ions. This is the central theoretical result of the work we present here. In this formulation, we assume that the fundamental charge carriers are no longer the free ions as in equation \ref{eq:NE}, but the whole range of ionic clusters. In equation \ref{eq:cNE}, the approximation we make is that \textit{clusters do not interact with one another}: we therefore name this approach ``cluster Nernst-Einstein'' (cNE). One can also unambiguously define a transference number, here for the cation: 

\begin{equation}\label{eq:t_cNE}
t^{+}_{\text{cNE}}= \displaystyle\frac{\sum_{i=0}^{N_+}\sum_{j=0}^{N_-} i z_+ z_{ij} \alpha_{ij} D_{ij}}{\sum_{i=0}^{N_+}\sum_{j=0}^{N_-} z_{ij}^2 \alpha_{ij} D_{ij}}, 
\end{equation}

where $i$ is the number of cations in cluster $ij$. \\

We now test our newly developed method in an ideal case, in order to assess the upper bound of the error on the conductivity we can expect. In this toy model, we introduce three different types of particles, which we tag as solvent, anions, and cations. These particles only interact through a Lennard-Jones potential, there are no Coulombic interactions, although when estimating the ionic conductivity, we assume that cations and anions respectively carry a charge of $+1$ and $-1$. Furthermore, cations and anions only interact with the solvent particles, and not with themselves. In this way, we have a non-interacting model for the species of interest. Through the Lennard-Jones potential, we set the size of the anion to be larger than the cation (all parameters are reported in the Supplemental Material\footnote{See Supplemental Material at [url] for (1) the Lennard-Jones and harmonic bond parameters used within the toy model, (2) atomic charges of the ionic species in the \textsc{pcff+} model, (3) details on the determination of cluster's diffusion coefficients, and (4) a comparison with experimental conductivity data. }). Finally, we define clusters by adding a harmonic potential between selected pairs of ions; in doing so, we input $\left\langle\bm\alpha\right\rangle$. Therefore in this model, we have a set of non-interacting ionic clusters for which we know exactly the population. \\

\onecolumngrid
\begin{figure*}[t]
\centering
\includegraphics[scale=0.75]{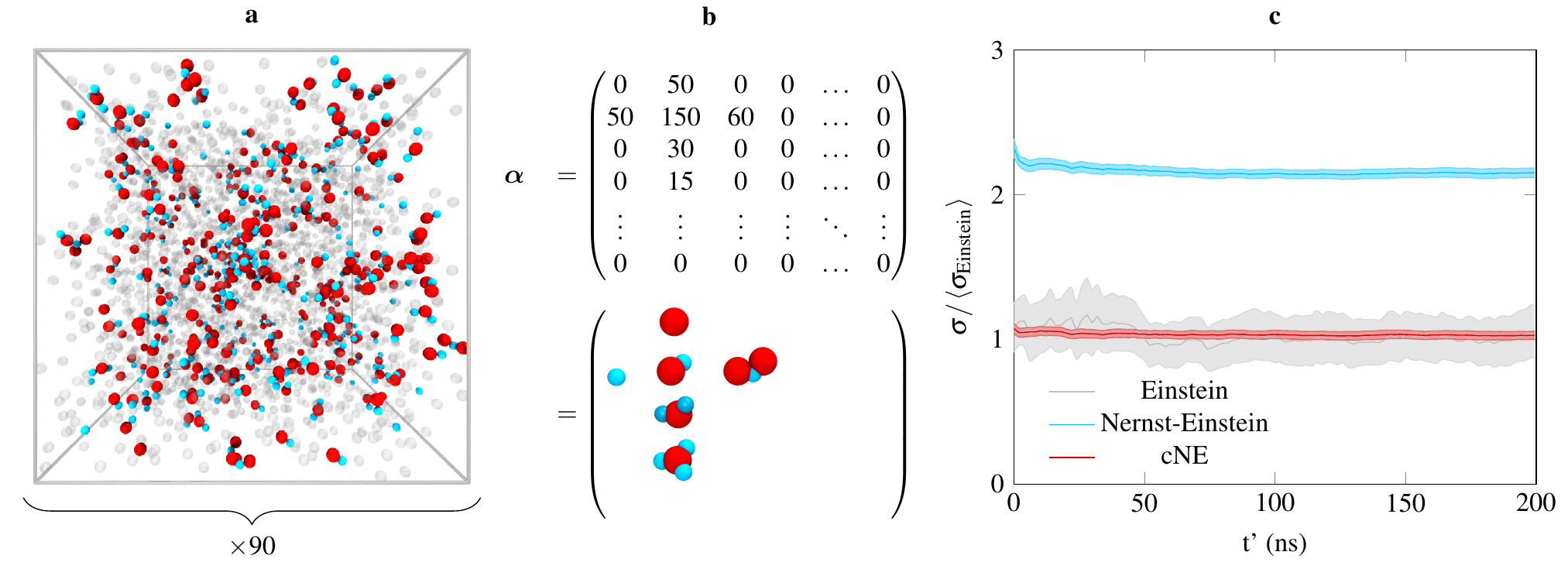}
\caption{Non-interacting toy model. (a) Atomistic model of one of the 90 independent configurations with anions, cations, and solvent particles respectively represented as red, cyan, and gray spheres, (b) cluster population, and (c) ionic conductivity over lag time obtained using the three different approaches. Conductivities are normalized using the average converged Einstein conductivity. }
\label{figure1}
\end{figure*}
\twocolumngrid

We equilibrate our system in the canonical ensemble at a temperature of 300 K, for 10 ns, using a timestep of 1 fs. Then, we sample atomic displacements for 200 ns in the microcanonical ensemble. We consider 90 independent configurations, constructed using a Monte-Carlo scheme. All of our results will only be averaged over these configurations, we do not perform any time-origin averaging. From the atomic displacements, we can compute $\bar{D}_+$, $\bar{D}_-$, as well as $\bm{D}$. From these quantities, we can estimate $\sigma$ using the Nernst-Einstein and the cNE approximations, which we will compare to the ``true'' Einstein conductivity. All results for this toy model are presented in figure \ref{figure1}. All molecular dynamics have been performed using the Large Atomic Molecular Massive Parallel Simulator (\textsc{lammps})\cite{plimpton1995} code, and all models prepared using the MedeA\textregistered\cite{MedeA} simulation environment. \\

We input on purpose a non-trivial cluster population. From figure \ref{figure1} (b), one can see that: (i) most of the ions are associated -- only less than 15 \% are uncoordinated -- and (ii) the cluster distribution is asymmetric. Conductivity results are presented in figure \ref{figure1} (c). All error bars reported in this Letter are 95 \% confidence intervals. As expected, the Nernst-Einstein approximation leads to an overestimation of the conductivity, by more than a factor of 2. The error on $\sigma_{\text{NE}}$ is however quite small ($\pm$ 1.4 \%), and convergence is reached rather fast. In contrast, the Einstein conductivity converges slowly, and the error remain rather large ($\pm$ 16.4 \%). Finally, the average converged cNE conductivity is in excellent agreement with the value obtained using the Einstein form ($1.03 \cdot \sigma_{\text{E}}$). The slight overestimation of the conductivity can be tentatively assigned to indirectly correlated motion (two clusters interacting with the same solvent particle will be indirectly correlated), but that is hard to confirm given the relatively large error on $\sigma_{\text{E}}$. The error on $\sigma_{\text{cNE}}$ is slightly higher than for the Nernst-Einstein case -- there are less particles to average on for $D_{ij}$'s compared to $\bar{D}_{\pm}$ -- but remains very small ($\pm$ 2.7 \%). Finally, the convergence is as fast as for the Nernst-Einstein approximation. \\ 

These results give us confidence that our approach can provide excellent estimates of the conductivity at a low computational cost. We next test our method on a more realistic system, a polymer electrolyte composed of poly(ethylene oxide) (PEO) and LiTFSI. There have been very recent experimental reports\cite{pesko2017} of negative cation transference numbers in such a system at high salt concentration, tentatively assigned to the existence of negatively charged ionic clusters. In addition to that, recent theoretical work\cite{molinari2018} showed that the cluster population is indeed asymmetric, with more anions than cations in clusters. Since PEO interacts strongly with Li cations, effectively there are less available cations to bind with anions, leading to this asymmetry. \\

\begin{figure*}[t]
\centering
\includegraphics[scale=0.75]{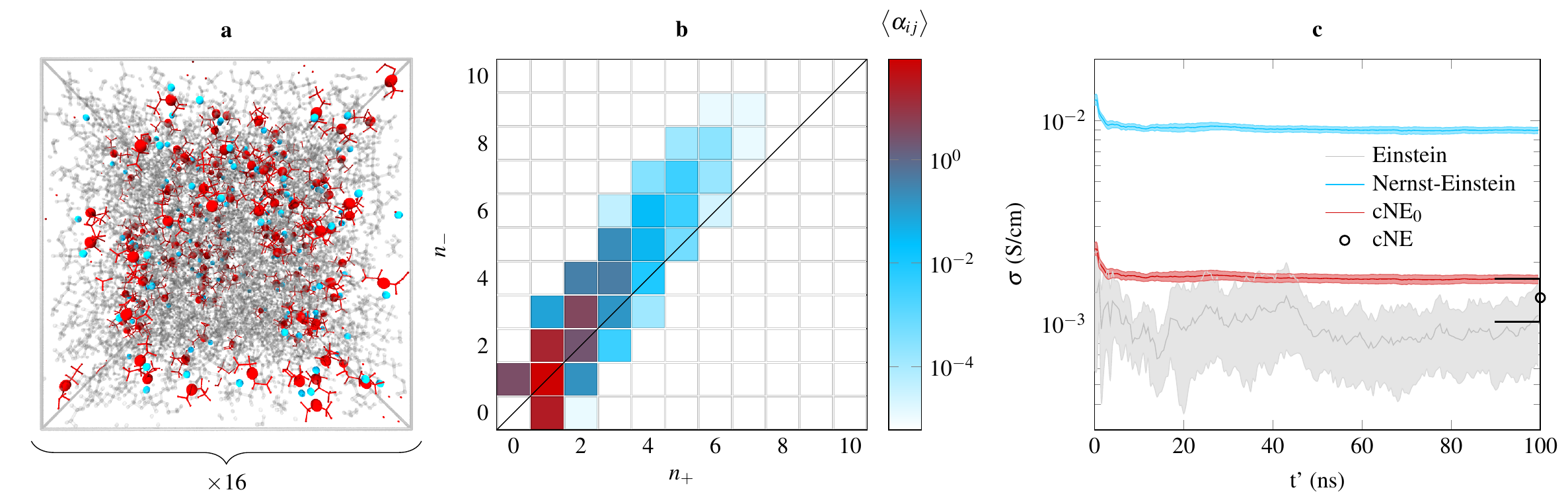}
\caption{PEO results. (a) Atomistic model of one of the 16 independent configurations. Atoms composing anions and cations are respectively represented as red and cyan spheres, and octaglyme atoms as gray spheres, (b) cluster population statistics (notice the logarithmic scale), and (c) ionic conductivity over lag time obtained through the Einstein, Nernst-Einstein, and cNE$_0$ approaches. The cNE conductivity is also shown. }
\label{figure2}
\end{figure*} 

We have performed molecular dynamics simulations on 16 independent realizations of a system composed of 200 (PEO)9 (also known as octaglyme) chains, with 150 LiTFSI molecules, leading to a salt concentration $r$ equal to 1/12 Li$^+$/EO. At such a concentration, experimentally reported transference numbers\cite{pesko2017} are still positive. However, the interatomic potential we use overestimates ionic association because of its point-charge model of electrostatics, as well as its absence of a description of polarization effects. This leads to a too ``hard'' model of the TFSI$^-$ anion. Therefore we expect our model to be representative of higher salt concentrations, for which negative cation transference numbers have been measured ($r \gtrsim$ 0.12 Li$^+$/EO). Interactions were modeled using the Polymer Consistent Force-Field (\textsc{pcff+})\cite{sun1994}, which has been previously employed to model PEO/LiTFSI electrolytes\cite{molinari2018}. Long-range Coulombic interactions have been accounted for using a particle-particle particle-mesh solver (\textsc{pppm}), and the Lennard-Jones term was truncated at 12 \AA. We have used the reversible reference system propagator algorithm (r\textsc{respa})\cite{tuckerman1992} to accelerate the simulations; topological contributions to the potential were integrated every 0.5 fs, while a time step of 2.0 fs was used for long-range electrostatics, and Lennard-Jones interactions. We have verified that the total energy was still correctly conserved in the microcanonical ensemble using these settings. The configurations were relaxed at a target temperature of 363 K and a pressure of 1 atm, using a series of energy minimizations and runs in the canonical and isothermal-isobaric (\textit{nPT}) ensembles\cite{molinari2018}, for a total of 5 ns. Once thermalization is achieved, the atomic displacements were sampled in the \textit{nPT} ensemble at the same thermodynamical conditions as mentioned before, for 100 ns -- amounting to a total of 1.6 $\mu$s of sampled dynamics. \\

Cluster populations were obtained using a similar approach to the one developed by Molinari \text{et al.}\cite{molinari2018}. We postulate that a cation and an anion form a cluster if the distance between the cation and any negatively charged atom of the anion is smaller than a certain cutoff value, set to 3.25 \AA. Importantly, ionic clusters do not have a unique definition. Using this approach, Molinari {et al.} noticed that when varying the cutoff distance, the main features of the cluster population which we will discuss later are conserved. \\

The cluster population is presented in figure \ref{figure2} (b). All errors on the matrix elements of $\left\langle \bm{\alpha} \right\rangle$ are on the order of $\pm$ 1 \%, and have been propagated in the evaluation of $\sigma_{\text{cNE}}$. The distribution calculated is indeed asymmetric, with overall more negatively charged clusters. 58 \% of all ions form pairs ($\alpha_{11}$). There are significantly more free cations ($\alpha_{10} \sim$ 27.5) than free anions ($\alpha_{01} \sim$ 3.1). Finally, a large number ($\alpha_{12} \sim 18.6$) of negatively charged triplets exist, as well as, to a lesser extent, negatively charged clusters involving five ions ($\alpha_{23} \sim$ 4.2). All of these observations are consistent with the assumption that due to the Li$^+$--EO interaction, there are less available cations to form clusters with anions. \\

Conductivities are reported in panel (c) of figure \ref{figure2}. The Nernst-Einstein and Einstein conductivities are plotted as a function of lag time. A convenient approximation to the cNE approach named cNE$_0$, where the diffusion coefficient of a given cluster is set as either $\bar{D}_+$ or $\bar{D}_-$ if it involves more cations or anions, is also plotted as a function of lag time. We could evaluate the diffusion coefficients of all of the clusters aforementioned since enough statistics were available, by directly tracking the cluster displacements. The higher order clusters with populations less than 0.5 were not taken into account in the cNE summation. Since clusters decay in 1--10 ns, we cannot plot $\sigma_{\text{cNE}}$ over lag time, which is why it is included on the right side of the plot as a single point. More details concerning the determination of diffusion coefficients are included in the Supplemental Material. The Nernst-Einstein approach spectacularly overestimates the conductivity by a factor of 10 ($\sigma_{\text{NE}} = 9.0 \cdot 10^{-3}$ S/cm, $\sigma_{\text{E}} = 8.8 \cdot 10^{-4}$ S/cm). However, the error on the measurement is an order of magnitude smaller than using the Einstein approach, respectively $\pm$ 3.6 \% and $\pm$ 40.0 \%, and its convergence is very fast (less than 20 ns). On the other hand, it is not clear if the Einstein conductivity is fully converged, even after 100 ns. This exemplifies the extremely large amount of data required to obtain a rigorous estimate of the ionic conductivity. The error on $\left\langle\alpha_{ij}\right\rangle$'s being small, the error on the cNE$_0$ conductivity is only $\pm$ 4.8 \%, on par with the accuracy of the Nernst-Einstein approach. More importantly, its value ($\sigma_{\text{cNE}_0} = 1.6 \cdot 10^{-3}$ S/cm) is in much better agreement with the Einstein conductivity, overestimating it by less than a factor of two. Finally, the cNE conductivity is even closer to the Einstein value ($\sigma_{\text{cNE}} = 1.3 \cdot 10^{-3}$ S/cm), with however an error of $\pm$ 23.8 \%, due to the increased difficulty to obtain good estimates of the diffusion coefficient matrix. This error is still smaller than the Einstein error. A comparison with reported experimental measurements\cite{teran2011} is included in the Supplemental Material. \\

In addition to obtaining better estimates of the ionic conductivity at a lower computational cost, the cNE approach also allows us to gain fundamental understanding in ion transport mechanisms. An interesting observation is that the diffusion coefficients of charged clusters (including free ions) are systematically smaller than neutral clusters -- in the absence of an applied voltage at least. For instance, the diffusion coefficient of the free cation $D_{10} = 7.2 \pm 1.2 \cdot 10^{-7}$ cm$^2$/s is smaller than the one of the ionic pair $D_{11} = 10.3 \pm 1.0 \cdot 10^{-7}$ cm$^2$/s, which is also true for higher order clusters. Negatively charged clusters (for instance, $D_{12} = 9.2 \pm 3.6 \cdot 10^{-7}$ cm$^2$/s) are also faster than positively charged clusters. This ranking can be understood in terms of interaction strength, with positively charged species strongly interacting with octaglyme, effectively slowing them down. This effect seems to dominate the scaling of diffusion rates. At higher cluster sizes, considering small charge numbers, we can speculate that the Stokes-Einstein behavior is recovered, with cluster size controlling diffusion rate. \\

Finally, we estimate the cation transference number. In the Nernst-Einstein framework, it is simply defined as a ratio between diffusion coefficients: 

\begin{equation}\label{eq:tNE}
t_{\text{NE}}^+ = \frac{\bar{D}_+}{\bar{D}_+ + \bar{D}_-}
\end{equation}

Because of the predominance of ion pairs, $\bar{D}_+$ and $\bar{D}_-$ are almost equal ($\bar{D}_+ = 9.8 \pm 0.4 \cdot 10^{-7}$ cm$^2$/s, $\bar{D}_- = 10.2 \pm 0.3 \cdot 10^{-7}$ cm$^2$/s), which leads to almost balanced contributions to the Nernst-Einstein conductivity: $t_{\text{NE}}^+ = 0.49 \pm 0.02$. Using equation \ref{eq:t_cNE}, we can estimate $t^+$ with the cNE and cNE$_0$ approximations. We obtain $t_{\text{cNE}_0}^+ = -0.04 \pm 0.05$, and $t_{\text{cNE}}^+ = -0.09 \pm 0.35$. While the uncertainty on these transference numbers is quite large due to error propagation, we still clearly calculate, for the first time to the best of our knowledge, negative values using molecular dynamics simulations. We confirm the hypothesis that a large population of negatively charged clusters involving cations can lead to negative transference numbers. To a lesser extent, the smaller diffusion coefficient of free Li$^+$ ions lowers even more $t^+$, as  can be seen by comparing the average values obtained using the cNE$_0$ and cNE approximations. These observations could be used as design rules for electrolytes: if the solvent molecules were to bind more strongly to the anion than to Li$^+$\cite{savoie2017}, the effect could be reversed, with more cations involved in clusters, and a distribution leaning towards positively charged aggregates. By doing so, we speculate that under the right conditions, $t^+ > 1$ could be obtained. \\

In summary, by considering ionic clusters as non-interacting species, we provide an intermediate approximation to the ionic conductivity -- more precise than the usual Nernst-Einstein equation -- both accurate and computationally inexpensive, and well suited to high salt concentrations. In addition, our method allows to understand physical mechanisms of ion transport in concentrated electrolytes. Finally, using the present approach, we are able to reproduce recent experimental results reporting negative cation transference numbers at high salt loading in PEO-based electrolytes, and to confirm the mechanism responsible for this phenomenon. \\


%
%


\bibliography{aps_bib}

%
%
%
%
%
%

\end{document}